\documentclass[a4paper,11pt]{article}

\usepackage{graphicx}
\usepackage{dcolumn}
\usepackage{bm}
\usepackage{amsmath}
\usepackage{amsfonts}

\newcommand{\dif}{{\rm d}}
\newcommand{\equal}{\buildrel {\rm def} \over {=} }

\newcommand{\vett}[1]{\mathbf{#1}}

\newcommand{\icsi}{x^{(i)}}
\newcommand{\yi}{y^{(i)}}
\newcommand{\pyi}{p_y^{(i)}}
\newcommand{\pxi}{p_x^{(i)}}

\title{Relaxation  properties in classical diamagnetism } 

\author{A. Carati\footnote{ Universit\`a di Milano, Dipartimento di Matematica
Via Saldini 50,  20133 Milano (Italy)  E-mail: {\tt
andrea.carati@unimi.it} } \and 
\addtocounter{footnote}{1}
F.~ Benfenati\footnote{ Universit\`a di
    Milano, Corso di Laurea in Fisica Via Celoria  12,  20133 Milano (Italy)}
 \and L. Galgani\footnote{ Universit\`a di Milano,
Dipartimento di Matematica, Via Saldini 50,  20133 Milano (Italy)  
E-mail: {\tt luigi.galgani@unimi.it} } }

\date{\today}

\begin{document}

\maketitle
\par

\noindent
Running title: classical diamagnetic properties of matter 
\vskip 3.truecm


\begin{abstract} 
In the present paper the problem of the relaxation of magnetization to
equilibrium (i.e., with no magnetization)
is investigated numerically for a variant of the well--known
model introduced by Bohr to study the diamagnetism of electrons in
metals. Such a model is mathematically equivalent to a billiard with
obstacles in a magnetic field.
We show that it is not guaranteed that equilibrium is
attained within the typical time scales of microscopic dynamics. 
Indeed, considering an out of equilibrium state produced by an
adiabatic  switching on of a magnetic field, we show that,
depending on the values
of the parameters, one has a relaxation  either  to equilibrium  or to a
diamagnetic (presumably metastable) state. 
The analogy with the relaxation properties in the FPU problem is also
pointed out.
\end{abstract}

\section{Introduction}

It is well known (see Refs. \cite{vleck,peierls}) that, according to classical
statistical mechanics, at equilibrium matter doesn't exhibit any
diamagnetic effects. For a classical model of free electrons in metals
this was first stated  
by Bohr (see \cite{bohr}). 
Indeed, he pointed out that the Maxwell--Boltzmann distribution of the
electron velocities depends, at a fixed temperature, only on the system's
energy.  On the other hand, the value of energy does not depend
on the magnetic field. So the electron velocity distribution, being
unaffected by the magnetic field, is the same as in the case of a
vanishing field, and thus  magnetization vanishes. The
same line of reasoning is followed in the general case; see for example 
 the classical textbook by Feynman \cite{feynman}.

On the other hand, a magnetization can exist in an out of equilibrium
state, as typically occurs when a magnetic field is switched on,
because this induces an electric field which does perform some work on
the system. In the very words of Bohr (see Ref. \cite{bohr},
page 382): ``\emph{It must be pointed out, however, that in a piece of
metal exposed to a variable magnetic field there will arise a
collective motion of the electrons -- the so called Foucault current
-- produced by the electric field which, according to the
electromagnetic theory, is inextricably connected with the variation
of the magnetic field; this current will produce a magnetic effect
that will counteract the variation of the magnetic
field\ldots}''. Bohr, however, also added ``\emph{\ldots the induced
  collective motion of the 
electrons will disappear very rapidly after the magnetic field has
become constant, without leaving any permanent effect upon the
statistical distribution of the velocities of the electrons}''. In
other words, a magnetization is induced by the switching on of the
field, but it will disappear when the system will have relaxed to the
final equilibrium.  In addition, Bohr guessed that the relaxation should always be
rapid. In the present paper we perform some numerical computations
in order to check whether this is actually the case. We will show that
there are cases in which, even in systems as billiards (which may be
very chaotic, see \cite{buminovich,buminovich2}), there appear several time
scales for the relaxation to equilibrium. In some cases the
equilibrium is not even attained within the available simulation time,
analogously to what occurs in FPU systems.

The existence of a 
relaxation time to equilibrium  for magnetization
 is of particular importance because, if one performs
measurements on a time scale shorter than that  needed to reach 
equilibrium,  an effective nonvanishing  magnetization will show up.
  In particular, if the system remains frozen in 
some metastable state, this diamagnetic property may be mistaken for a
true equilibrium one. In any case, the system will then actually
present some ``effective'' diamagnetic property, notwithstanding the
fact that magnetization vanishes at equilibrium.  The situation is
reminiscent of  that of  FPU systems (see Refs. \cite{cgg,fpu,cggp}),
where the lack of 
thermalization leads to values of the specific heat smaller than the
equilibrium one.  So the question is the following one: if initially (when
the magnetic field   vanishes) the system is distributed according the to Gibbs
statistics, and then a magnetic field is ``slowly'' increased up to a final
value $\vett B_0$, how much time is it  necessary for the system to
reach a state in which the magnetization vanishes again?

The problem of the rate of thermalization for magnetization will be
dealt with in this paper through numerical computations, making reference
to a variant of the well--known Bohr model for free electrons in metals,
which will be described in a moment.  We will consider a system of $N$
electrons and compute numerically their trajectories, so that the
total magnetization (which, we recall, is proportional to
the total angular momentum) can be evaluated at any time.  

The macroscopic magnetization should in principle be defined through 
a Gibbs average over the initial data in phase space. However, in the present
paper  we will take a more pragmatic approach, and will define the
macroscopic magnetization through a ``moving average'' of 
the ``instantaneous'' total magnetization over a fixed time interval
for  a single orbit, as will be explained more precisely later.
The aim is to study the behaviour of  magnetization as time increases,
and in particular to see whether it vanishes, or rather it settles
down to some constant value.

We first consider the original Bohr model, i.e., a set of noninteracting
electrons moving on a plane normal to the magnetic field $\vett B$,
and confined in a domain of a simple shape (in particular we consider
circular domains). The numerical simulations show that, after the
field has been switched on,  in this model
there is no relaxation at all, so that one ends up with a perpetual 
nonvanishing magnetization.

Thus, we modify the model and insert inside the domain some fixed
obstacles simulating the interaction of the electrons with the ionic
lattice of a metal, so that we end up with a dynamical system which
is just a billiard in a magnetic field. Systems of this type are well
studied in the literature (see for example 
\cite{biglia1, biglia2, biglia3, biglia4}), with the aim of
determining how do their ergodic properties depend on the magnetic 
field.  At variance with the quoted papers, however, we deal here with a
time--dependent magnetic field, just because we are interested in studying the
effects of an adiabatic switching on of the field, which produces a
nonequilibrium state of magnetization.
We find that, in the modified Bohr model, 
after the field was switched on indeed in general one can have
relaxation to equilibrium or to a metaequilibrium state according to
the values of the parameters.

The paper is organized as follows. In Section~2 we define the model,
and briefly discuss the numerical method for integrating the equations
of motion. We also check that, in the case of a constant $\vett B$,
the magnetization vanishes, in agreement with Bohr's theorem.
In Section~3 we discuss the case in which the magnetic field $\vett B$
is switched on adiabatically, and illustrate the main result of the
paper, namely that situations of apparent metastability may occur. 
In Section~4 the results are discussed. 
In an Appendix we recall how, according to Linear Response Theory, 
the diamagnetic susceptibility is related to the time--autocorrelation
of  magnetization. 

\section{The model}

The model concerns a system of $N$ identical
point particles of mass $m$ and charge $q$, moving in a plane. We
denote by $\vett x_i=(\icsi,\yi)$, $i=1,\ldots,N$, the coordinates of
the $i$--th particle and by $\vett p_i=(\pxi,\pyi)$ their conjugate
momenta. The magnetic field is taken perpendicular to the
plane and homogeneous, i.e., one has $\vett B=(0,0,B(t))$, so that
the vector potential $\vett A$ at  point $\vett r$ is given by
$\vett A(\vett r) = \frac 12 \vett B \wedge \vett r$, where $\wedge$
denotes the vector product. The particles do not interact with each
other, and the Hamiltonian is simply given by
\begin{equation}\label{eq:ham}
H(\vett p_i,\vett x_i,t) = \frac 1{2m} \sum_{i=1}^{N} \Big( \vett p_i
    - \frac {q}{2c} \vett B(t) \wedge \vett x_i\Big)^2 +
    \sum_{i=1}^{N} V(\vett x_i) \ ,
\end{equation}
where $c$ is the speed of light, while $V(\vett r)$ is a confining
potential, i.e. a function vanishing inside the allowed domain and
diverging outside it (corresponding to a boundary condition of elastic
reflection).  We take for $m$ and $q$ the mass and the charge
of the electron, and we use atomic units, in which the electron mass,
the electron charge and the reduced Planck constant $\hbar$ are all set equal
to 1.  The number $N$ of particles in most simulations is taken in
the range $10^3 \div 10^4$, and in some cases is increased up to
$10^5$.

The time dependence of the magnetic field is taken as 
\begin{equation}
\vett B(t) = \frac {\vett B_0}2 \Big( 1+\tanh \frac {t-t_i}{t_c} \Big)
   \ ,
\end{equation} 
where $t_c$ is the characteristic time over which the magnetic field
varies, and $t_i$ is, in a sense, the time at which the field is switched
on. Indeed, for times $t-t_i<- 5t_c$ the magnetic field essentially vanishes,
while for $t-t_i>5t_c$ the field is essentially constant, equal to
$\vett B_0$.
The values of the magnetic field $B_0=|\vett B_0|$ are taken in the
range $10^{-3} \div 10^{-2}$ (in atomic units), while $t_c$ is taken of
order $10^6$ (in our time unit) and $t_i$ of the order
$5\;10^7$.  
For what concerns the domain, we take it of two different kinds: the
first one is a ``simple domain'', i.e. a circular domain of radius
$R$, with $R$ in the range $5\times 10^3 \div 5\times 10^4$
(expressed in atomic units, i.e. the Bohr radius). We will show in the
next Section that for this type of domain the model shows no
relaxation at all.

The second type of domain is chosen in order to describe a more
realistic situation, in which the electrons interact with the ionic
lattice of a metal. Thus, we add inside the circular domain a square
lattice of circular obstacles. The radius $R$ of the circle is taken
fixed equal to $5\times 10^4$, while the radius $r$ of each obstacle
is taken equal to $10$, and the lattice step in the range from $10^3$
to $10^4$.

From Hamiltonian (\ref{eq:ham}) one gets the equations of motion,
which are all decoupled because the particles have no mutual
interaction.  Such equations are numerically integrated using the
``leap frog'' algorithm\footnote{In some cases we checked the
algorithm using a symplectic method of higher order, and we found that
the results agree (within the numerical errors) with the simpler leap
frog method.} which we now briefly describe.  The integration step of
the ``leap frog'' method from the values $\vett p_i,\vett x_i$ at time
$t$ to the values $\vett p'_i,\vett x'_i$ at time $t+\tau$ is
performed by a canonical transformation with generating function
\begin{equation}
   S = \sum_{i=1}^{N} \vett p_i \cdot \vett x'_i - \tau H(\vett
     p_i,\vett x'_i,t+\tau) \ .
\end{equation}
This gives an implicit integration scheme, which can be easily made
explicit for the particular form of our Hamiltonian (\ref{eq:ham}).

The collision with the boundary is dealt with in the following way:
one performs an integration step and then checks whether the particle
remains inside the allowed domain. If this is not the case, the
evolution is made not according to the leap frog algorithm, but as
follows: the position $\vett x_i$ and the component of $\vett
p_i$ tangent to the boundary are left unchanged, while the normal
component of the momentum $\vett p_i$ changes its sign.  One checks
that this approximation is sufficient to keep constant the energy of
the system.

Finally, the initial data are taken at random in the following way:
the initial positions of the particles are uniformly distributed
inside the allowed domain, while the initial velocities are
distributed according to a Maxwell--Boltzmann distribution at a
temperature $T$. We take a temperature $T$ 
such that $k_BT=1/250$ in our units ($k_B$ being the
Boltzmann constant). This temperature
corresponds to a mean velocity  approximately equal to
$1/3000$ the speed of light, which is a typical velocity of the
electrons in metals at ordinary conditions (see \cite{tdt}).

We now explain how the magnetization $\vett M$ is computed.  The
instantaneous value of  magnetization, let us call it $\mathfrak{
M}$, is given by
\begin{equation}\label{eq:mag}
\mathfrak{M} = \frac q{2c} \sum_{i=1}^{N} \vett {x}_i \wedge \dot
      {\vett x}_i \equiv \frac q{2mc} \vett L_{tot} \ ,
\end{equation}
where $\vett L_{tot}$ is the total angular momentum. Such a quantity
is found to exhibit, in our numerical computations, 
some fluctuations. So, in
order to smooth them out, we perform  a moving average, i.e.,
we  report, in place of the
instantaneous values, the corresponding time averages over a small
time interval (of the order of one hundredth  of the total
integration time).

\begin{figure}[ht]
  \begin{center}
    \includegraphics[width=0.5\textwidth]{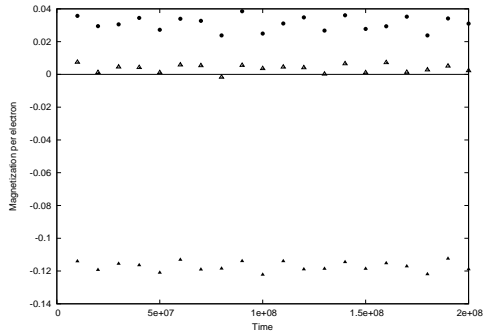}
  \end{center}
\caption{\label{fig:fig1} Plot of the component of $\vett{M}(t)$ 
  along the field $\vett B$ versus time in the original
  Bohr model,   for three different samples of 10$^3$ electrons. The
  magnetic field is equal to 0.001.} 
\end{figure}

Some examples of the results of this procedure are shown in Figure~1,
which refers to the case $|\vett B|=0.001$ independent of time, and to three
different samples of $N=1000$ electrons each. At first sight it
appears that for each sample the magnetization is constant and different from zero. This
actually is a statistical effect due to the finiteness of the system
(1000 electrons).
Indeed, as pointed out by Bohr, the value of  magnetization turns
out to be due to the
contributions of two populations.  The first one is the population of
the bulk electrons, which stay deep inside the domain without hitting
the boundary. They move in circles, producing a magnetic field which is
directed against the external one.

The second population is made of electrons which repeatedly hit the
boundary and produce a current  directed opposite to that of the
first population. The two contributions would exactly cancel each other
in the limit of an infinite number of electrons. This was the
conclusion of Bohr.

In our simulations the populations are finite, so that the two
contributions are slightly different.  To evaluate this statistical
effect we perform ten different simulations (with $B=0.01$ and
$N=10^5$); in five cases we take into account the contribution of the
boundary, while in the remaining ones we eliminate the boundary,
i.e. we just put $V\equal 0$ into
Hamiltonian (\ref{eq:ham}).  The mean and the standard deviation (over the
samples) of the magnetization per electron $|\vett M|/N$ for the cases cases are
reported in Table~\ref{tab:1}. 
\begin{table}[ht]
  \caption{\label{tab:1} Magnetization per electron in the case of a
    constant field (without obstacles), in the presence or in the
    absence of the boundary. The magnetic field is taken
    equal to 0.01, the radius of the circle is taken equal to $5000$, the
    number of electrons is taken equal to 10$^5$.  }
  \begin{center}
    \begin{tabular}{|l|c|c|}
      \hline ~ & mean & standard deviation \\ \hline with boundary & -0.0007 &
      0.0024 \\ without boundary & -0.3994 & 0.0022 \\ \hline
    \end{tabular}
  \end{center}
\end{table}
As one sees, in the absence of the
boundary the magnetization per electron turns out to be finite. 
This fact, by the way, was already pointed out\footnote{ Actually, 
  such a paper discusses the problem  in the frame of stochastic processes
  (Langevin or Fokker Planck equations), rather than in the frame of a
  deterministic Hamiltonian  dynamics, as we do here. However,  the
  physical problem addressed is essentially the same. } in a
recent paper (see \cite{kumar}). Instead, when  the boundary is present
(first line of Table~\ref{tab:1}), the   
results are consistent with a vanishing magnetization, as predicted by Bohr.

Things are however different if one considers a time--dependent
magnetic field, as occurs when the field is  adiabatic switched on. 
This is discussed  in the next Section.

\section{The case of a time--dependent field}

\begin{figure}[th]
  \begin{center}
    \includegraphics[width=0.5\textwidth]{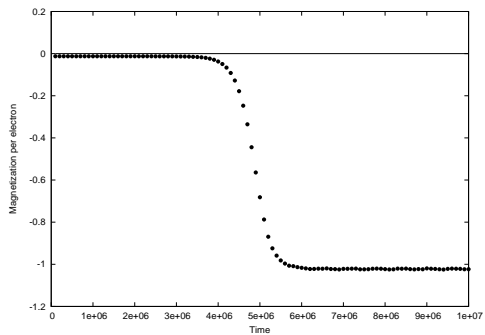}
  \end{center}
\caption{\label{fig:fig2} Plot of the component of $\vett{M}(t)$ 
  along the field $\vett B$ versus time, for a
  circular domain of radius 5000 and no obstacles, with a
  magnetic field adiabatically switched on with a characteristic time
  equal to $5\times10^6$.  The number of electrons is 5000, and the final
  field strength is 0.01.}
\end{figure}

\begin{figure}[thp]
  \begin{center}
    \includegraphics[width=0.3\textwidth]{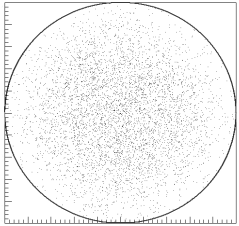}
  \end{center}    
  \caption{\label{fig:fig4} Plot of the positions of the electrons at
    the end of the simulation considered in Figure~\ref{fig:fig2}. }
\end{figure}

The first result in the presence of a time--dependent field is
summarized in Figure~\ref{fig:fig2},
which refers to the original Bohr model, i.e., the case of a circular
domain without obstacles. We see that in this case there is no
relaxation at all to a vanishing magnetization, so that the model
exhibits a fully diamagnetic behaviour.  The reason is apparent from
inspection of Figure~\ref{fig:fig4}, in which the positions of the
electrons at the end of the same simulation considered in
Figure~\ref{fig:fig2}, are shown. Indeed, one sees
that the electrons are no more uniformly distributed in the available
domain, but are instead concentrated away from the boundary.  This
corresponds to the appearance of a magnetic pressure which confines
the electrons away from the boundary, analogously to what occurs in
the magnetic confinement in Tokamaks.


Thus, the contribution to magnetization due to the ``bulk''
electrons and that due to those hitting the boundary 
become different, the Bohr compensation does not occur anymore, and
the magnetization turns out to be different from 
zero. This shows that, in order to obtain a classical
diamagnetic effect, it is not 
necessary to physically eliminate the border from the model (as was
done in Ref.~\cite{kumar} for a case of a constant field),
because a diamagnetic effect already manifests itself even in the
presence of the border, just
as  a consequence of the adiabatic switching on of the field.

From a mathematical point of view, the appearance of a
magnetization is due to the fact that, in the case of a circular
domain, the total angular momentum is conserved once the magnetic
field becomes static.

\begin{figure}
  \begin{center}
    \includegraphics[width=0.5\textwidth]{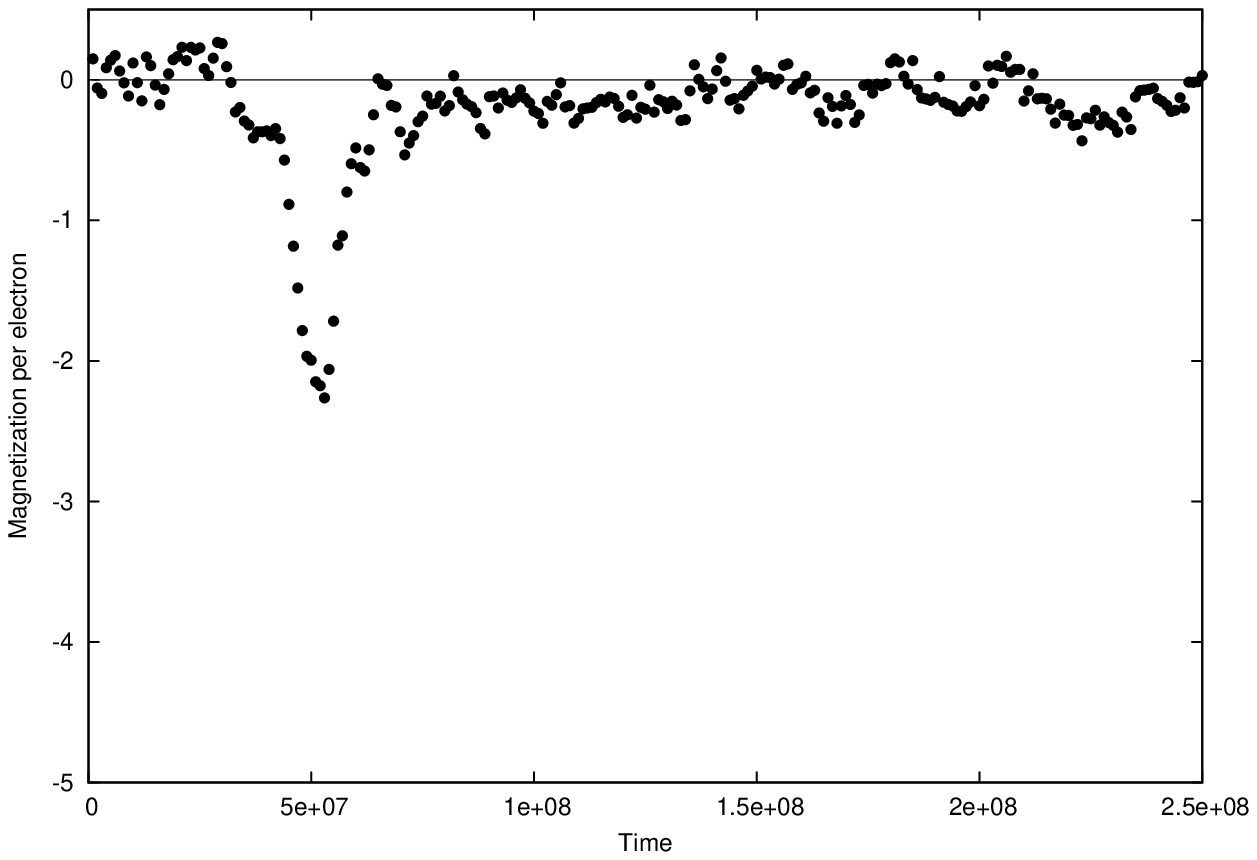}
  \end{center}

  \begin{center}
    \includegraphics[width=0.5\textwidth]{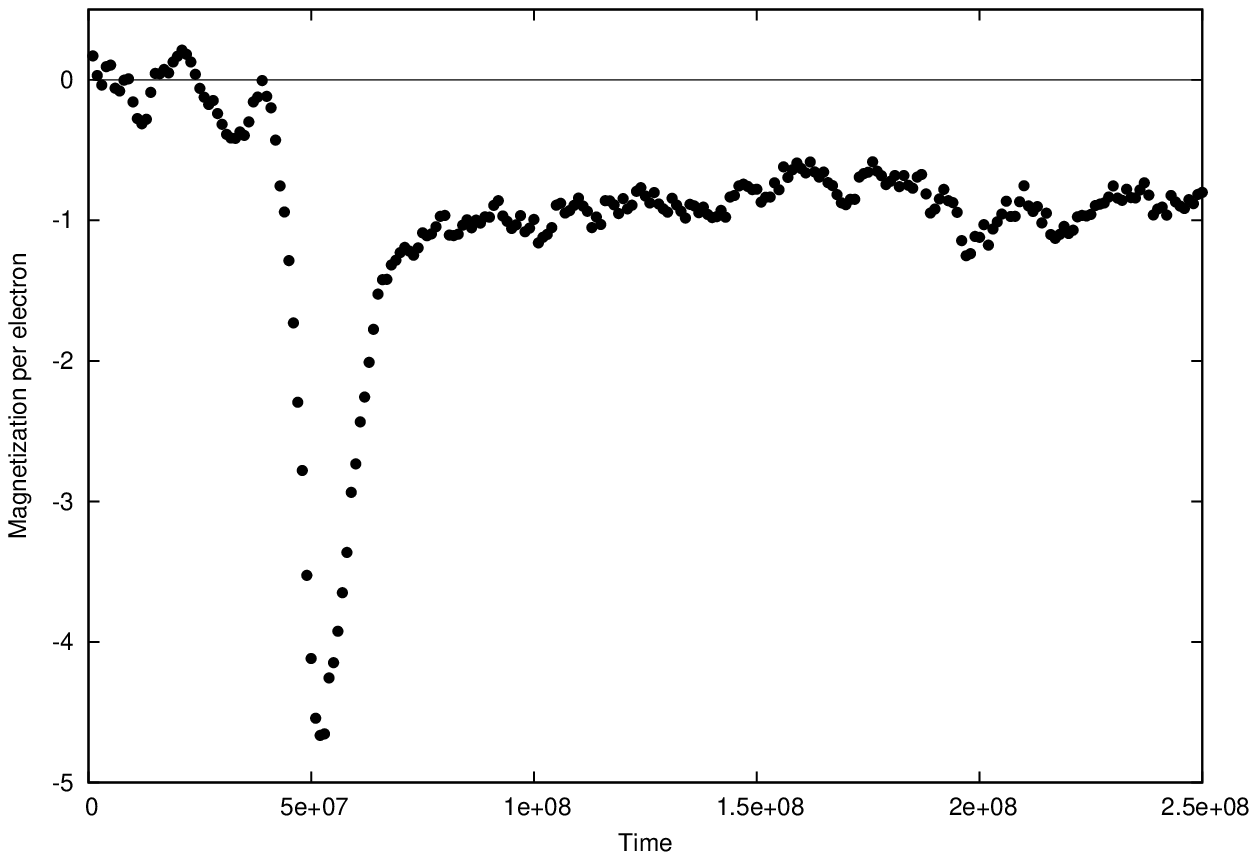}
  \end{center}
  \caption{\label{fig:fig6} Plot of the component of $\vett{M}(t)$ 
    along the field $\vett B$ versus time in the
    case of a circular domains and a lattice of obstacles. The field
    strength is equal to 
    $0.001$, the switching time $t_c$ is equal to $5\times 10^6$ and the
    number of electrons is $N=2500$. Upper
    panel:  distance between obstacles equal to 2500. Lower panel:
    distance between obstacles equal to 4500.}
\end{figure}
However, in a more realistic model of electron conduction in a metal,
the total angular momentum of the 
electrons alone would not be conserved, because of the scattering of the
electrons by the ions (electron--phonon interaction), although the
total angular momentum is obviously still conserved. In order to mimic
the ionic lattice of a metal,  
we change the shape of the domain, by inserting a square
lattice of circular obstacles. Thus, the  total angular momentum of the
electrons is no more conserved. We first  checked numerically
that this occurs, as it should,  even  in the
absence of a magnetic field: starting from a positive value, the total
angular momentum was found to  steadily  decrease towards zero.  The speed of
convergence, however, depends on the parameters of the model (number of
electrons, distance between the obstacles, and so on). 

Then, we consider the case of interest, in which a field is
adiabatically switched on, and the results are summarized in
Figure~\ref{fig:fig6}. The upper panel refers to a distance between
the obstacles equal to $2.5\times 10^3$ (with $B_0= 0.001$, $t_c=5
\times 10^6$ and $N=2500$). Here it is clearly exhibited that the
system relaxes to a non diamagnetic state in a time of order
$10^7$. The behaviour, however, changes if the parameters are
varied. For example if one increases the distance between the
obstacles up to $4.5\times 10^3$ (less than by a factor two) while keeping
the other parameters fixed, the resulting plot of magnetization versus
time is given in the lower panel.  One sees that the relaxation occurs
with (at least) two time scales: after a fast relaxation to a
nonvanishing value (with a time scale of the same order of magnitude
as in the previous case), there follows a much slower relaxation which
we actually were unable to follow up to the end.

\section{Conclusions}

In the present paper we pointed out the role of chaoticity of dynamics
in providing or not the relaxation of magnetization to a vanishing
value, i.e., to equilibrium. Indeed, if the dynamics is not chaotic,
as in the original Bohr model corresponding to a circular billiard
without obstacles, there is no relaxation at all. Instead, if the
system is sufficiently chaotic, as is the modified Bohr model
corresponding to a circular billiard with obstacles, then the
magnetization decays to zero, as required at equilibrium.  Actually,
this was found to occur for suitable values of the physical parameters
of the modified Bohr model, whereas for different values of the
parameters  the magnetization does not decay to zero
within the available time. In the latter case an effective
diamagnetism shows up, in the vein of the metastability phenomena
which are familiar for example in the frame of glasses and also in the
FPU problem. This possibly is the main result of the present paper.

This fact may have some physical significance. Indeed, in the
literature there are reported evidences of empirical metastability
phenomena for the magnetic susceptibility. See for example Ref.
\cite{wills} (see also Ref. \cite{willis2}), in which a
hysteresis curve is shown for
the diamagnetic constant of water. See also Ref. \cite{hysteresis} for 
diamagnetic hysteresis in beryllium, and Ref. \cite{supercon}
for constricted diamagnetic hysteresis loops in high 
critical temperature superconductors.


As a final comment we recall that, in the spirit of Linear Response
Theory, the relaxation to equilibrium of magnetization can also be
discussed, through the Fluctuation--Dissipation theorem, in terms of
the decaying to zero of the time--autocorrelations of magnetization
itself.  Indeed, one has
\begin{equation} \label{eq:mag4}
\langle \vett M(t)\rangle = \beta \int_{t_0}^t\dif s\, \dot{\vett
  B}(s)\cdot \langle \vett M(s) \vett M(t)\rangle \ ,
\end{equation}
as is briefly recalled in the Appendix. Thus, essentially, one has to
study the time--autocorrelation of 
angular momentum in a billiard with obstacles in a magnetic field. By
the way, one can also just consider the case of constant magnetic
field, with a suitable initial out of equilibrium state of magnetization. For
the time--correlation in billiard flows in the absence of a magnetic
field, see Ref. \cite{chernov}.

\appendix

\section*{Appendix: diamagnetism by the Green--Kubo relations}

We deduce  here the expression,  given in the
Conclusions, for the magnetization of a body according to the
Fluctuation--Dissipation theorem. Consider the Hamiltonian
(\ref{eq:ham}) which we rewrite here,
slightly changing the notation, as
\begin{equation} 
H = \sum_{j=1}^n \left[ \frac 1{2m}\Big(\vett p_j - \frac {e}{2c}\vett B
\wedge \vett q_j \Big)^2 +V(\vett q_j) \right] \ .
\end{equation} 
where the magnetic field $\vett B(t)$ depends on time explicitly, being
switched on adiabatically from zero up to its final value.  Notice
that now, at variance with the main text, the electron charge is denoted by
$e$, while $\vett q_j$ denotes the position of the $j$--th electron.
Then, the Gibbs distribution
\begin{equation}
\rho_0 = \frac {e^{-\beta H_0}}{Z_0(\beta)} \ ,
\end{equation} 
 where $H_0$ is the
Hamiltonian evaluated at zero field, will  only be a zero--th order
approximation of the true distribution 
$\rho$. This, we recall, has to satisfy the Liouville equation
\begin{equation}\label{eq:Liou}
  \dot \rho + [H,\rho] = 0 \ ,
\end{equation}
(where $[~,~]$ denotes  Poisson bracket), together with the
asymptotic condition $\rho \to \rho_0$ for $t\to -\infty$ (because the
distribution should coincide with the Gibbs one before the magnetic field is
turned on). Suppose now that the magnetic field can be treated as a
small parameter, and expand the distribution $\rho$ in powers of it. 
Setting\footnote{ We dispense for a moment with the normalization
constant $Z(\beta)$, which is known to be  independent of
time, and thus equal to $Z_0(\beta)$. }  $\rho = e^{-\beta H}(1 +
\rho_1 + \ldots) $, 
and substituting it into the Liouville equation, one gets for $\rho_1$ the
equation
\begin{equation}\label{eq:Liou1}
  \dot \rho_1 + [H,\rho_1] = - \beta\, \dot{\vett B}\cdot \vett M \ ,
\end{equation}
where the magnetization $\vett M$ is given by
\begin{equation}\label{eq:magn1}
  \vett M \equal \frac e{2c} \sum \vett q_j\wedge \dot{\vett q}_j =
  \frac e{2mc} \sum \vett q_j \wedge \Big(\vett p_j - \frac {e}{2c}\vett B \wedge \vett
  q_j \Big) \ .
\end{equation}
We note that the magnetization, as a dynamical variable, depends
explicitly on time besides on the point $\vett x$ of phase space,
so that we will write sometimes $\vett M=\vett M(\vett x,t)$ in order to
emphasize this fact.
Denoting by $\Phi_{t_0}^t$ the flow associated with  Hamilton's
equations at time $t$ with initial data taken at time $t_0$ (remember
that Hamilton's equations are not autonomous, so that it is mandatory to
specify the time at which initial data are taken), if one looks for a
solution of (\ref{eq:Liou1}) in the form $\rho_1(\vett
x,t)=\chi(\Phi_{t_0}^t \vett y,t)$, then $\chi$ has to satisfy
\begin{equation}
\partial_t \chi(\Phi_{t_0}^t \vett y,t) = - \beta \, \dot {\vett
  B}\cdot \vett M(\Phi^t_{t_0}\vett y,t)
\end{equation}
The function $\chi$ is thus simply obtained by integration, so that,
putting $\vett x=\Phi^{t_0}_t \vett y$ in the resulting
expression, one finds
\begin{equation}\label{eq:rho1}
  \rho_1(\vett x,t) = - \beta \, \int_{t_0}^t \dif s \, \dot {\vett
           B}(s) \cdot \vett M(\Phi^s_t\vett x,s) \ .
\end{equation}
Here we used the group property $\Phi^s_{t_0} \Phi^{t_0}_t =
\Phi^s_t$ of the flow; furthermore, the lower integration limit $t_0$ is
intended to be a time before the magnetic field is switched on.


Recall now that the
normalization constant, being time independent, 
is nothing but the  partition function $Z_0(\beta)$
computed for a  vanishing  field, because it can be
computed at time $t_0$, i.e. when the magnetic field
vanishes. Thus, 
to first order in the magnetic field, the magnetization at time
$t$ is given by
\begin{equation}  
    \langle \vett M(t) \rangle = \beta \int_{t_0}^t \dif s \, \dot
      {\vett B}(s) \cdot \int_{\mathcal{M}}\dif \vett x \, \frac
      {e^{-\beta H(\vett x,t)}}{Z_0(\beta)} \vett M(\Phi^s_t\vett x,s)
      \vett M(\vett x,t) 
\end{equation}
i.e. by
\begin{equation}
\langle \vett M(t) \rangle  = \beta \int_{t_0}^t \dif s \, \dot
      {\vett B}(s) \cdot \langle \vett M(s) \vett M(t) \rangle \ , 
\end{equation}
where $\vett M(s)$ is the magnetization evolved backwards in time up to
time $s$, starting from data at time $t$, and the averages are
performed with respect to the Gibbs distribution at time $t$. 
This is the formula  given in the Conclusions. In this
expression the average is performed with respect to the \emph{final
data}. An analogous expression could also
be given with the average performed with respect to the
\emph{initial data}, but we do not insist here on this point.

\end{document}